\documentclass[12pt,preprint]{aastex}
\def\beq{\begin{equation}}
\def\eeq{\end{equation}}
\def\ben{\begin{eqnarray}} 
\def\een{\end{eqnarray}}
\def\lcdm{\Lambda{\rm CDM}}
\def\dunit{\,h^{-1}\,{\rm Mpc}}
\def\munit{\,h^{-1}\,M_{\odot}}
\def\mth{M_{\rm th}}
\def\msth{m_{{\rm th},\star}}
\def\bj{\hat{\bf j}}
\def\bja{\hat{\bf j}_{a}}
\def\bjb{\hat{\bf j}_{b}}

\def\lms{\log\left(M_{\star}/[h^{-1}\,M_{\odot}]\right)}

\def\bd{\hat{\bf d}}
\def\inc{\vartheta_{\rm I}}
\def\vid{\rm VIDE}
\def\rev{\rm REVOLVER}
\def\vf{\rm VoidFinder}
\usepackage{xcolor}

\begin{document}
\title{An Observed Transition of Galaxy Spins on the Void Surfaces}
\author{Jounghun Lee$^{1}$ and Jun-Sung Moon$^{1,2}$}
\affil{$^1$Department of Physics and Astronomy, Seoul National University, Seoul 08826, Republic of Korea
\email{cosmos.hun@gmail.com,jsmoon.astro@gmail.com}}
\affil{$^2$Research Institute of Basic Sciences, Seoul National University, Seoul 08826, Republic of Korea}
\begin{abstract}
In the linear theory, the galaxy angular momentum vectors which originate from the initial tidal interactions with surrounding matter distribution 
intrinsically develop perpendicular alignments with the directions of maximum matter compression, regardless of galaxy mass. 
In simulations, however, the galaxy spins exhibit parallel alignments in the mass-range
lower than a certain threshold, which depends on redshift, web type, and background cosmology.  
We show that the observed three dimensional spins of the spiral galaxies located on the void surfaces from the Sloan Digital Sky 
Survey indeed transit from the perpendicular to the parallel alignments with the directions toward the nearest void centers at the threshold zone, 
$9.51\le\log [M_{th,\star}/(\munit)]\le10.03$.  This study presents a first direct observational evidence for the occurrence of the mass-dependent spin 
transition of the real galaxies with respect to the non-filamentary structures of the cosmic web, opening a way to constrain the initial conditions 
of the early universe by measuring the spin transition threshold. 
\end{abstract}
\section{Introduction}\label{sec:intro}

What is conventionally denominated the {\it galaxy spin transition} is a phenomenon that the angular momentum vectors (spins) of low-mass galaxies 
exhibit parallel alignments with the directions of maximum compression of surrounding matter, while those of high-mass counterparts manifest perpendicular 
alignments. The occurrence of the galaxy spin transition was witnessed by numerous N-body simulations 
\citep[e.g.,][]{ara-etal07,hah-etal07,paz-etal08,cod-etal12,tro-etal13,AY14,dub-etal14,for-etal14,cod-etal15a,cod-etal18,gan-etal18,
wan-etal18,gan-etal19,kra-etal20,lee-etal20}, 
which in turn propelled a flurry of theoretical and observational exertions to find a physical explanation and a practical evidence for it 
\citep{lib-etal13,tem-etal13,TL13,cod-etal15b,zha-etal15,pah-etal16,WK17,kro-etal19,wel-etal20,bar-etal22,tud-etal22,ML23}.

In spite of these endeavors, no established theoretical model for its mechanism nor highly significant detection of the galaxy spin transition from 
observations have so far been made. 
On the theoretical side, the main difficulty is to coherently explain the complex behavior of the threshold mass, $\mth$, at which the galaxy 
spin transition occurs. The value of $\mth$ was shown by numerical experiments to sensitively depend on such various factors as redshift, 
environment, scale, web type and background cosmology \citep{ara-etal07,hah-etal07,cod-etal12,tro-etal13,zha-etal15,cod-etal18,lee-etal20,LL20,gan-etal21}. 
The conventional scenario which attributed the galaxy spin transition to the effect of hierarchical merging along anisotropic cosmic web 
\citep{ara-etal07,hah-etal07,pic-etal11,cod-etal12,tro-etal13,dub-etal14, cod-etal18,gan-etal18,kro-etal19} failed to provide a coherent physical explanation 
for the variation of $\mth$ with these factors.  Furthermore, this conventional scenario was shown to be inconsistent with the recent numerical finding 
of \citet{LM22} that the alignment strengths of the galaxy spins with the directions of maximum matter compression do not show any strong variation 
with the latest merging epochs. 

On the observational side, the most challenging task is to measure the galaxy spins ($\bj$) and directions of maximal matter compression 
($\bd$) as accurately as possible in practice. 
Although several authors reported possible observational detections of the mass dependent spin transition of the real galaxies, 
the low statistical significance of their signals as well as relatively low accuracy in the measurements of $\{\bj, \bd\}$ beleaguered their 
results with cautions. 
For example, \citet{wel-etal20} claimed a first direct detection of the mass dependent spin transition of the field galaxies with respect to the 
nearby filaments from the Sydney–AAO Multi-object Integral-field spectrograph (SAMI) survey \citep{sami}, under the assumption that 
$\bd$ is perpendicular to the elongated filament axes. Their analysis, however, relied on the simple approximation of the three dimensional 
spins of the SAMI galaxies by the position angles alone as well as on the measurements of the filament axes in the two dimensional 
projected space.

\citet{bar-etal22} analyzed the same SAMI data but with an improved methodology to detect a signal of the mass dependent spin transition of 
the galaxies. They improved the measurement of the galaxy spins by employing the scheme proposed by \citet{LE07} which basically reckons 
the shape of a galaxy as a circular thin disk, and properly took into account the two-fold ambiguity that the sign of the radial component of a 
galaxy spin along the line-of-sight direction cannot be determined by the scheme based on the circular thin disk approximation 
\citep{lee11,kra-etal21}.  They also improved the measurement of $\bd$ by efficiently eliminating the Finger-of-God (FoG) 
effect \citep{fog} caused by the peculiar motions of group/cluster galaxies embedded in the filaments \citep[see also][]{kra-etal18}. 

\citet{bar-etal22} assumed that the spins of the elliptical galaxies should be in the direction of their minor shape axes and thus could be obtained 
by the same scheme applied to the spiral galaxies. Yet, recent N-body simulations disproved this assumption, 
demonstrating that the galaxy spins are not perfectly aligned with their minor shape axes and that the shape axes do not show any mass-dependent 
spin transition \citep{LM22}, which implies that the scheme of \citet{LE07} is invalid for the measurements of the elliptical galaxy spins. 
Since the $\bj$-$\bd$ alignments are expected to be quite a weak signal vulnerable to systematic contaminations, it is essentially important 
to diminish all known systematics to the lowest levels for a detection of the galaxy spin transition and for the measurement of $\mth$. 

Besides, the previous observational works determined $\bd$ as a direction perpendicular to a thin straight filament nearest to a given 
galaxy. However, the real filaments in the universe have much more complex shapes, whose geometrical properties have been found to 
severely affect the strengths and tendencies of the alignments with the galaxy spins \citep[see][and references therein]{gan-etal21}. 
In other words, the systematic errors in the measurement of $\bd$ are likely to be produced not only by the FoG effect but also by the deviation of the 
filament shapes from thin straight lines. Given these simple approximations and assumptions made in the previous works, it was not a surprise 
that the solidity of their claimed detections have been protested by several counter-evidences against the occurrence of the galaxy spin transitions 
\citep[e.g.,][]{zha-etal15,che-etal19,kro-etal19,tud-etal22}. Henceforth, it is still inconclusive whether or not the mass-dependent galaxy spin 
transition indeed occurs in reality. 

Here, we suggest that the cosmic voids, large empty regions devoid of galaxies in the universe, be the most optimal web-type from which a signal 
of the occurrence of the galaxy spin transition could be detectable with high significance.  The logic behind this suggestion can be found in the 
context of the linear tidal torque theory \citep{whi84,lp00,lp01}. In this theory, the protogalaxies acquire angular momentum at first order and 
develop preferential spin alignments with the principal axes of the local tidal tensors, only if the principal axes of their inertia momentum tensor are not 
perfectly aligned with those of the local tidal tensor. 
It turned out, however, that the principal axes of the two tensors are quite strongly aligned with one another with alignment strengths sensitively dependent 
on the environmental density \citep{lee06}.   The lower density the environment has, the less strong alignment the principal axes of the two tensors exhibit.  
Therefore, it may be more probable to find a statistically significant signal of the spin alignment and its transition with respect to the principal axes of 
the local tidal field from the galaxies located in the neighborhood of cosmic voids -- the web-type with lowest environmental densities. 

For the galaxies at the void outskirts, $\bd$ can be determined as a direction toward the void centers without making any simplified assumption about the
 void shapes nor concerning about the FoG effect. Analyzing the most updated catalogs of the observed 
voids, we attempt here to explore how the three dimensional spins of the spiral galaxies are aligned with the directions toward the nearest void centers and 
to investigate if a significant signal of the mass-dependent galaxy spin transition can be found. We will focus only on the spiral galaxies whose 
three dimensional spins can be rather accurately determined by the circular thin disk approximation up to the aforementioned sign ambiguity. 
Throughout this paper, we will assume a flat universe with dominant cosmological constant ($\Lambda$) and cold dark matter (CDM).  

\section{Physical Analysis}\label{sec:anal}

From the Seventh Data Release of the Sloan Digital Sky Survey \citep[SDSS DR7,][]{sdssdr7}, \citet{void-catalog} produced 
three different catalogs of cosmic voids in the redshift range of $0\le z\le 0.114$ for the Planck and and the Wilkinson
Microwave Anisotropy Probe 5 year (WMAP5) $\lcdm$ cosmologies. The three catalogs correspond to three different void-identification 
algorithms, namely,  the VIDE \citep{vide}, REVOLVER \citep{revolver}, and VoidFinder \citep{HV02}. 
The former two implement the same ZOBOV scheme \citep{zobov} based on the Voronoi tessellation density estimator, 
but differ in the pruning steps through which the linked zones of local density minima are selectively divided into individual voids 
\citep{vide,revolver}. 
Meanwhile, the VoidFinder finds the voids as merged volumes of maximum empty spheres in the spatial galaxy distributions \citep{HV02}. 
For more information on the void finding process, we refer the readers to \citet{void-catalog} and references therein. 
A total of $531$, $518$ and $1163$ ($535$, $519$ and $1184$) voids were identified via the $\vid$, $\rev$, and $\vf$ algorithms, respectively, 
from a volumed limited sample of the SDSS DR7 galaxies for the planck (WMAP5) cosmology, whose median radii turned out to be 
in the range of $15\le R_{\rm v, med}/(\dunit)\le 19$ \citep{void-catalog}. 

To be statistically consistent with the void catalogs of \citet{void-catalog}, we utilize a spectroscopic galaxy catalog from the SDSS DR7
compiled by \citet{hue-etal11}, which contains information on how probable it is for a given galaxy to be classified as a certain 
morphological type. To each galaxy in the catalog is assigned four different probabilities, $P(E)$, $P(S0)$, $P(Sab)$, and $P(Scd)$, 
of its morphology being elliptical, lenticular, early type spiral, and late-type spiral, respectively. 
To the galaxy catalog of \citet{hue-etal11}, we apply the following three cuts for the selection of only {\it large spiral galaxies}: 
$\Gamma_{s}\ge 20$ (size-cut), $P(Sab)\ge 0.5$ $\cup$ $P(Scd)\ge 0.5$ (morphology cut), and $z\le 0.11$ (redshift cut), 
where $\Gamma_{s}$ denotes the number of pixels that the apparent size of a given galaxy occupies in the SDSS frame. 
As we are going to use information on the isophotal quantities like position angles ($\beta)$ and axial ratios ($q$) of the selected spirals 
for the determination of their spins, we exclude those galaxies whose isophotal images are not large enough to guarantee reliable 
measurements of $\beta$ and $q$ \citep{SM16}. 

For each of the selected spiral galaxies, we determine its spin, $\bj$, from information on $\beta$ and $q$, with the help of the scheme devised by \citet{lee11} 
based on the circular thin disk approximation \citep[see also][]{LE07}. We first estimate its inclination angle, $\inc$, as 
\begin{equation}
\label{eqn:xi}
\cos^{2}\inc = \frac{q^{2}-p^{2}}{1-p^{2}}\, ,
\end{equation}
where $p$ is the "intrinsic flatness parameter" introduced by \citet{HG84} to take into account non-negligible thickness of a spiral disk due to the presence of a central bulge.  
To the $Scd$ and $Sab$ galaxies are conventionally assigned the $p=0.1$ and $p=0.2$, respectively \citep{HG84}.  
If the spiral arms of a given spiral galaxy viewed face-on are spinning in a clockwise manner, then $\bj$ can be determined in the spherical-polar coordinate system as 
\begin{eqnarray}
\label{eqn:jr}
\hat{j}_{r}&=&\cos\inc\, , \\
\label{eqn:jt}
\hat{j}_{\theta}&=&(1-\cos^{2}\inc)^{1/2}\sin\beta\, ,\\ 
\label{eqn:jp}
\hat{j}_{\phi}&=&(1-\cos^{2}\inc)^{1/2}\cos\beta\, .
\end{eqnarray}

Since it is usually very hard in practice to determine whether a spiral galaxy is spinning in a clock-wise or counterclock-wise manner, 
the true sign of a galaxy spin along the line of sight direction cannot be readily determined, so called the two-fold ambiguity of the spin 
measurement \citep{lee11}. Accepting this sign ambiguity, we will consider both of the possibilities, i.e., clockwise and counter clockwise spinning as in \citet{lee11}.
Rotating the frame to the equatorial Cartesian system, we end up having two different spin realizations, $\bja=(\hat{j}_{ai})$ and $\bjb=(\hat{j}_{bi})$,  
for each selected spiral galaxy as
\begin{eqnarray}
\hat{j}_{a1}&=&\hat{j}_{r}\sin\theta\cos\phi + \hat{j}_{\theta}\cos\theta\cos\phi  - 
\hat{j}_{\phi}\sin\phi ,\\
\hat{j}_{a2} &=&\hat{j}_{r}\sin\theta\sin\phi  + \hat{j}_{\theta}\cos\theta\sin\phi + 
\hat{j}_{\phi}\cos\phi ,\\
\hat{j}_{a3} &=& \hat{j}_{r}\cos\theta - \hat{j}_{\theta}\sin\theta , 
\end{eqnarray}
\begin{eqnarray}
\hat{j}_{b1}&=&-\hat{j}_{r}\sin\theta\cos\phi + \hat{j}_{\theta}\cos\theta\cos\phi  - 
\hat{j}_{\phi}\sin\phi ,\\
\hat{j}_{b2} &=&-\hat{j}_{r}\sin\theta\sin\phi  + \hat{j}_{\theta}\cos\theta\sin\phi + 
\hat{j}_{\phi}\cos\phi ,\\
\hat{j}_{b3} &=& -\hat{j}_{r}\cos\theta - \hat{j}_{\theta}\sin\theta , 
\end{eqnarray}
where $\theta=\pi/2-\delta_{I}$ and $\phi=\alpha$ with declination $\delta_{I}$  and right ascension $\alpha$. 

\begin{figure}[ht]
\centering
\includegraphics[height=15cm,width=16cm]{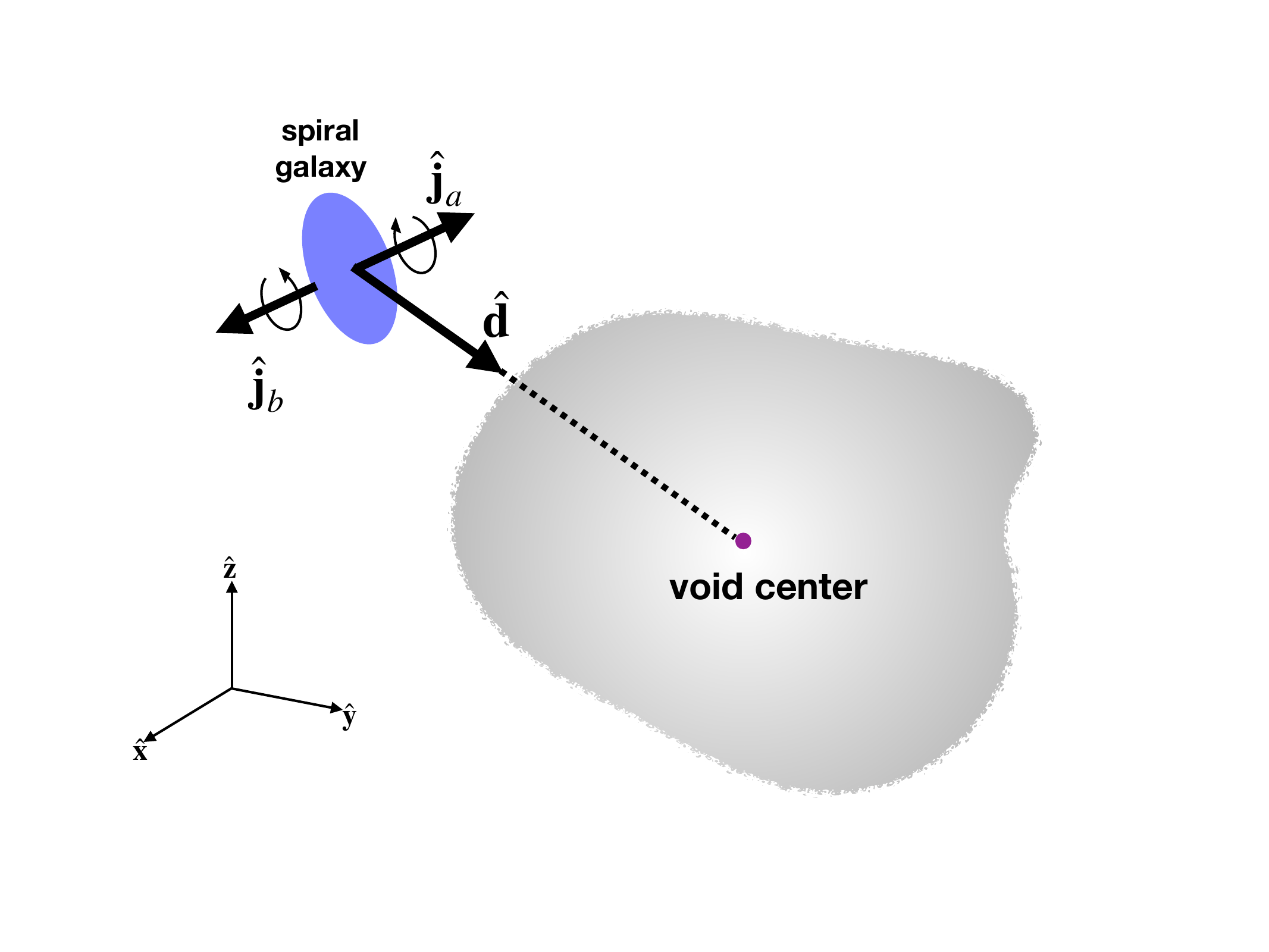}
\caption{Schematic representation of two different realizations of the 3D spins of a spiral galaxy and the direction 
toward the nearest void center.} 
\label{fig:schematic}
\end{figure}

Using information on the equatorial Cartesian coordinates of each selected spiral galaxy, we first measure its separation distances from all of 
the voids to search for the nearest one.  The unit direction vector, $\bd$, from a given spiral galaxy located at ${\bf x}_{g}$ toward 
the center of its nearest void,  ${\bf x}_{v}$, is calculated as $\bd \equiv \left({\bf x}_{g}-{\bf x}_{v}\right)/\vert{\bf x}_{g}-{\bf x}_{v}\vert$. 
Then, two realizations of the alignment between the spiral galaxy spin and the direction to its nearest void are computed as $\vert\bja\cdot\bd\vert$ 
and $\vert\bjb\cdot\bd\vert$. Figure \ref{fig:schematic} illustrates a schematic view of $\bja$, $\bjb$ and $\bd$ for a SDSS spiral galaxy in the neighborhood 
of a cosmic void. 

Obtaining information on the stellar masses, $M_{\star}$, of the selected spiral galaxies from the estimates made by \citet{men-etal14} with the help of 
the broadband spectral energy distributions,  we also apply two additional cuts to the selected spirals: $\vert{\bf x}_{g}-{\bf x}_{v}\vert \ge R_{\rm v}$ 
and $m_{\star}\equiv\lms >0$, where $R_{\rm v}$ denotes the effective radius of a nearest void. The first condition is required to exclude those few spirals located 
inside the nearest voids, while the second condition excludes those spirals for which no information on the stellar mass is available. 
The selected spirals are then split into six samples of equal sizes according to $m_{\star}$ ($\{S_{1}\}_{i=1}^{6}$ in a  $m_{\star}$-increasing order). 
The first six rows of Table \ref{tab:ng} list the numbers of the selected spiral galaxies, $N_{g}$, belonging to $\{S_{1}\}_{i=1}^{6}$ for the six 
different cases (three different void-identification algorithms and two different cosmologies).  
The dependence of $N_{g}$ on the void-identification algorithm comes from the fact that the numbers of the excluded galaxies depend on 
the void abundance. 

The ensemble average of the alignments between the galaxy spins and the directions to the nearest void centers is taken separately over each of  $\{S_{1}\}_{i=1}^{6}$ as
\beq
\label{eqn:ali}
\langle\vert\bj\cdot\bd\vert\rangle=\frac{1}{2N_{g}}\left(\sum_{\gamma=1}^{N_{g}}\vert\hat{\bf j}_{a,\gamma}\cdot\bd_{\gamma}\vert + \sum_{i=1}^{N_{g}}\vert\hat{\bf j}_{b,\gamma}\cdot\bd_{\gamma}\vert\right)\, ,
\eeq
where $\hat{\bf j}_{a,\gamma}$ and $\hat{\bf j}_{b,\gamma}$ are the two spin realizations of the $\gamma$th spiral galaxy whose unit direction toward its nearest 
void is denoted by $\bd_{\gamma}$. The errors in $\langle\vert\bj\cdot\bd\vert\rangle$ are obtained with the help of the bootstrap method. 
A significantly higher (lower) value of $\langle\vert\bj\cdot\bd\vert\rangle$ than $0.5$ will in principle signal a parallel (perpendicular) alignment of the spiral galaxy 
spins with the directions to the nearest void centers, where the reference value of $0.5$ comes from the theoretical expectation of the ensemble average in case of no 
$\bj$-$\bd$ alignments under the assumption that there is no systematic errors, either.

\begin{figure}[ht]
\centering
\includegraphics[height=18cm,width=14cm]{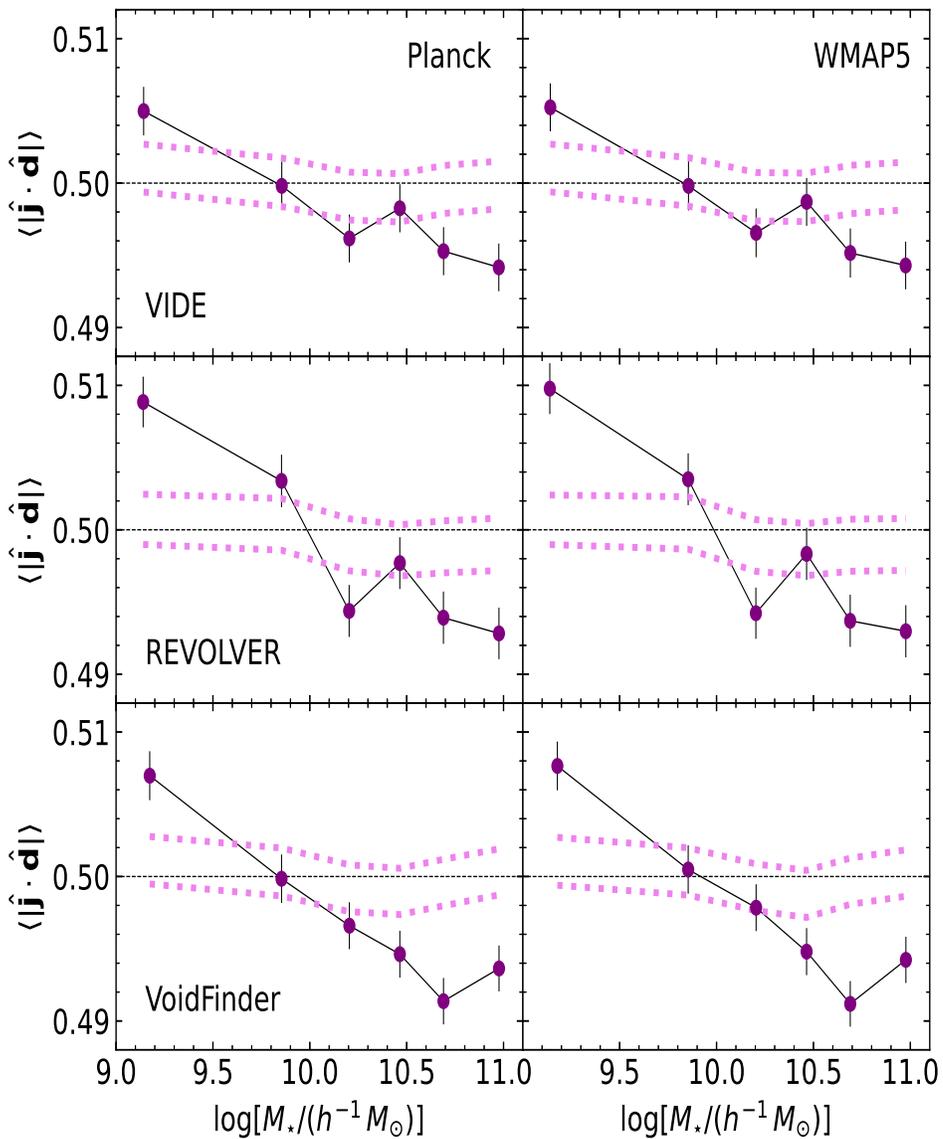}
\caption{Mean alignments between the galaxy spins and the directions toward the nearest void centers (purple colored filled circles) 
with bootstrap errors, along with $\pm 1\sigma$ scatters among the $1000$ randomly shuffled resamples (violet colored dotted lines). 
The main result is a detection of the occurrence of the stellar spin transition of the spiral galaxies on the void surfaces for the VoidFinder case.}
\label{fig:align}
\end{figure}

In reality, however, some unknown systematics could always exist, which would deviate the reference value from $0.5$ even in case of no true alignments 
between $\bj$ and $\bd$. 
To statistically deal with this unknown systematics, we create $10,000$ resamples by repeatedly and randomly shuffling the positions of the selected spiral galaxies but 
with fixing their spins. Then, we redo the whole analysis with each of the resamples to determine one standard deviation scatters ($1\sigma$) of $\langle\vert\bj\cdot\bd\vert\rangle$ 
averaged over the $10,000$ resamples \citep[see][and references therein]{TL13,kra-etal21,bar-etal22}. 
Figure \ref{fig:align} plots $\langle\vert\bj\cdot\bd\vert\rangle$ from the original samples (purple colored filled circles) with the bootstrap errorbars 
along with the $1\sigma$ scatters among the $10,000$ resamples (violet colored dotted lines), for the six different cases. 
As can be clearly seen, for all of the cases, the $\langle\vert\bj\cdot\bd\vert\rangle$ values show an overall trend of diminishing with the increment of $m_{\star}$ from above 
$0.5$ down to below  $0.5$. But only for the $\vf$ case, we note that both of the parallel and perpendicular spin alignments found at the lowest and highest 
mass bins (corresponding to $S_{1}$ and $S_{6}$, respectively), are statistically significant ($>3\sigma$), which implies the occurrence of the spin transition somewhere 
in between mass ranges.

To rigorously determine the threshold stellar mass range, $\Delta\msth$ (spin transition zone), at which the transition from parallel to perpendicular 
spin alignments really occurs, we employ a statistical methodology similar to the one suggested by \citet{LL20}. Basically, this methodology first sets up
a null hypothesis, $H_{\rm null}$, of no $\bj$-$\bd$ alignment at a given mass bin and performs a KS test of $H_{\rm null}$. For a given mass bin to be 
determined as $\Delta\msth$, the following two conditions must be satisfied: First, the KS test rejects $H_{\rm null}$ at the confidence level, 
${\rm CL}<99.9\%$, in the given mass bin. Second, the KS test rejects $H_{\rm null}$ at ${\rm CL}\ge 99.9\%$ in the adjacent mass bins 
both higher and lower than the given one. 

Under $H_{\rm null}$, the probability density of $\vert\bj\cdot\bd\vert$ is statistically equivalent to a uniform distribution provided that systematic contamination is negligible. 
For each of the six mass-selected sample, we compute the maximum difference between $P(\le\vert\bj\cdot\bd\vert)$ and $\vert\bj\cdot\bd\vert$, which 
are the cumulative versions of the probability density of $\vert\bj\cdot\bd\vert$ and the uniform probability density corresponding to $H_{\rm null}$, 
respectively. Then, we evaluate $\tilde{D}_{\rm max}\equiv (2N_{g})^{-1/2}D_{\rm max}$ whose value is expected to be higher (lower) 
than the critical value of $1.949$ \citep{LL20} if the KS test rejects $H_{\rm null}$ at ${\rm CL}\ge 99.9\%$ (${\rm CL}< 99.9\%$). 
If the spin transition occurs in a certain mass bin, we expect $\tilde{D}_{\rm max}<1.949$ from the corresponding sample but $\tilde{D}_{\rm max}>1.949$ from the other samples. 

\begin{figure}[ht]
\centering
\includegraphics[height=18cm,width=14cm]{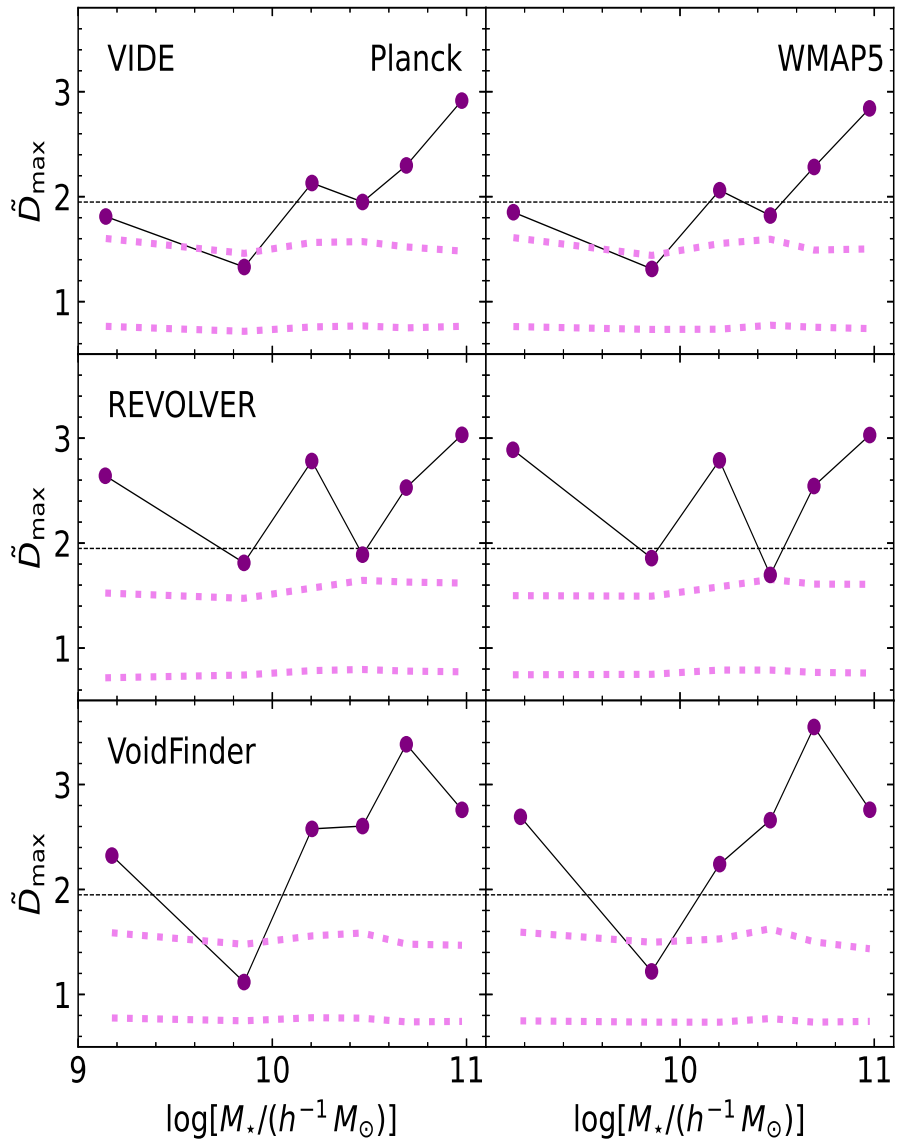}
\caption{Confidence levels (purple colored filled circles) at which the null hypothesis of no alignments is rejected by the KS one sample test, 
along with the $\pm 1\sigma$ scatters among the $1000$ randomly shuffled resamples (violet colored dotted lines). 
The main result is a constraint of the spin transition mass threshold to the range of $9.51\le\log\left[M_{\rm th,\star}/(\munit)\right]\le10.3$.}
\label{fig:cl}
\end{figure}

We also apply this method repeatedly to each of the $10,000$ resamples created by the random shuffling, and calculate $1\sigma$ scatters from the average $\tilde{D}_{\rm max}$ 
over the resamples. As can be seen in Figure \ref{fig:cl},  the randomly shuffled resamples yield almost constant values of $\tilde{D}_{\rm max}$ lower than $1.949$ 
within $1\sigma$ in the whole mass range (violet colored dotted lines). This result implies that the probability density function of $\vert\bj\cdot\bd\vert$ from the randomly 
shuffled resamples is statistically equivalent to the uniform probability density and thus that no signal of the spin transition can be found from the resamples. 
Meanwhile, the $\tilde{D}_{\rm max}$ values from the original samples (purple colored filled circles) exhibit fluctuating behaviors, higher or lower than $1.949$, 
depending on $m_{\star}$. For the VoidFinder case, we find $\tilde{D}_{\rm max}<1.949$ only from $S_{2}$ while the other samples yield $\tilde{D}_{\rm max}>1.949$, 
which implies the occurrence of the spin transition in the mass range of $S_{2}$, i.e., $\Delta\msth=[9.51,\ 10.03]$. 
Whereas, the VIDE and REVOLVER cases yield $\tilde{D}_{\rm max}<1.949$ from two non-adjacent samples, failing in satisfying the aforementioned two conditions. 

\begin{figure}[ht]
\centering
\includegraphics[height=18cm,width=14cm]{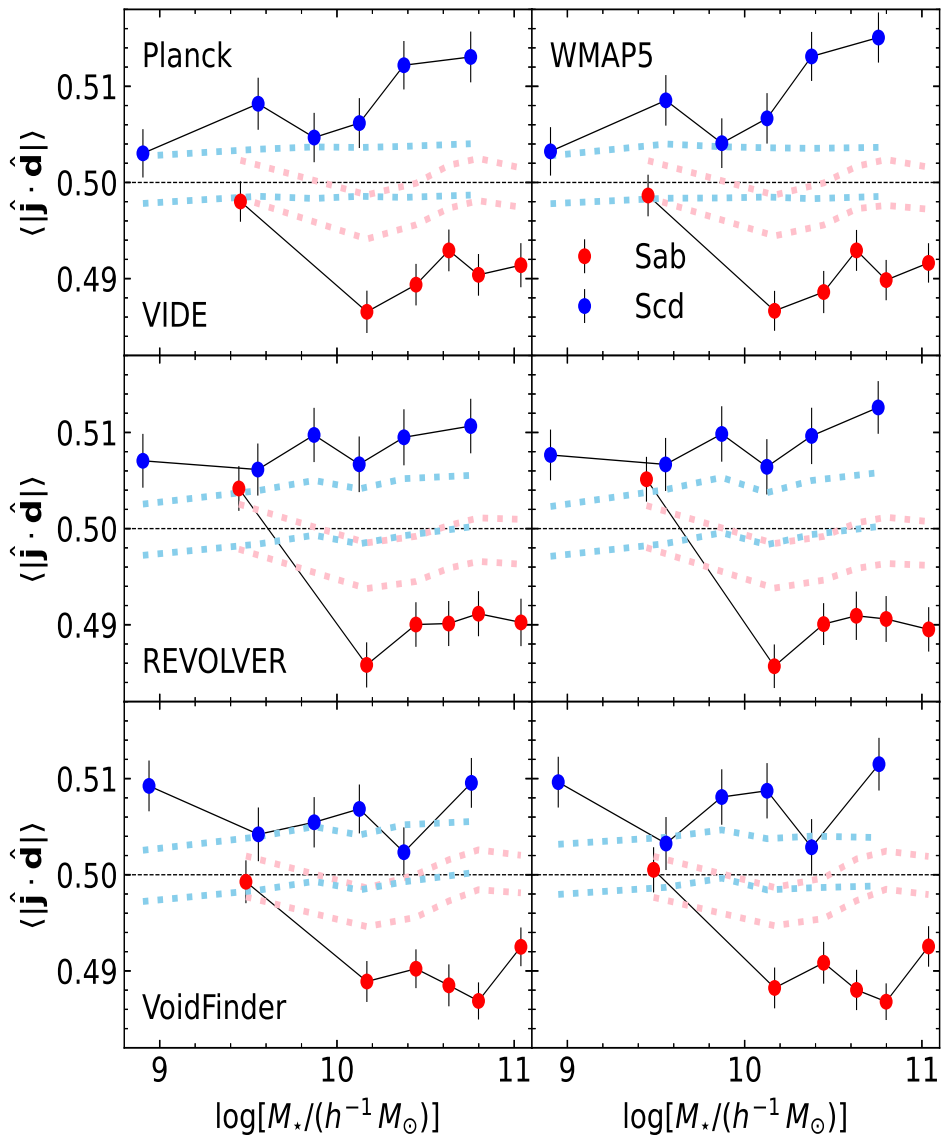}
\caption{Same as Figure \ref{fig:align} but with treating the $Sab$ and $Scd$ galaxies separately (red and blue colored 
filled circles, respectively), along with the $1\sigma$ scatters among the randomly shuffled resamples (pink and 
sky blue colored dotted lines). 
No signal of the occurrence of the stellar spin transition at a fixed morphological type.}
\label{fig:sep}
\end{figure}

We also redo the whole analysis but at a fixed morphological type by separately treating the $Scd$ and $Sab$ galaxies, whose numbers are 
listed in Table \ref{tab:ng}.  Figure \ref{fig:sep} shows $\vert\bj\cdot\bd\vert$ with Bootstrap errorbars from the original six samples of the $Scd$ and $Sab$ galaxies 
(blue and red colored filled circles, respectively) along with the $1\sigma$ scatters from the average over the randomly shuffled resamples of the two types 
(sky blue and pink colored dotted lines, respectively). 
Note that each of the six $m_{\star}$ selected samples composed of the $Sab$ galaxies has a different mass range
from that of the $Scd$ ones, although the same notations, $\{S\}_{i=1}^{6}$, are used for both of the cases. 
As can be seen, neither $Sab$ nor $Scd$ galaxy samples show a spin transition behavior for any of the six cases. From the $Sab$ galaxies, we find only 
perpendicular alignments between $\bj$ and $\bd$ in the almost entire mass range except for in the lowest mass bin corresponding to $S_{1}$ from which no signal of any 
alignment is found. In contrast, the $Scd$ galaxies yield only parallel alignments in the entire mass range, albeit statistically significant only in the high mass bins 
(corresponding to $S_{5}$ and $S_{6}$). 

\begin{figure}[ht]
\centering
\includegraphics[height=18cm,width=14cm]{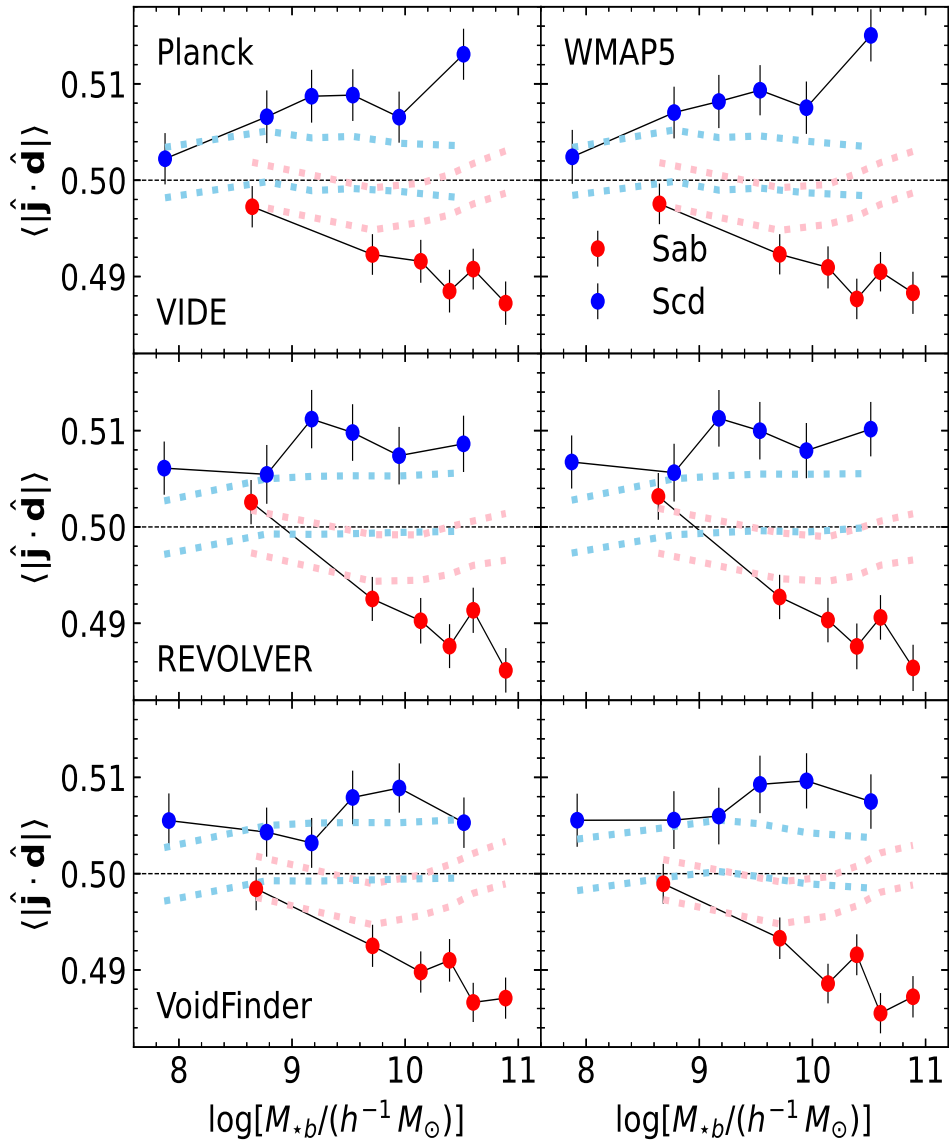}
\caption{Same as Figure \ref{fig:sep} but as a function of the bulge stellar mass. 
Disappearance of the spin transition signal at fixed morphology should not be due to the difference between 
the $Sab$ and $Scd$ galaxies in their bulge masses.}
\label{fig:bsep}
\end{figure}

Recalling the claim of \citet{bar-etal22} that the bulge mass is most strongly linked with the occurrence of the galaxy spin transitions 
as well as the well known fact that the $Scd$ galaxies have much lower bulge masses than the $Sab$ counterparts at a fixed total stellar mass, 
we recalculate $\langle\vert\bj\cdot\bd\vert\rangle$ and $1\sigma$ scatters from the $Sab$ and $Scd$ galaxies separately as a function of 
their bulge masses, $M_{\star b}$, which are also obtained from the estimates of \citet{men-etal14}. 
Figure \ref{fig:bsep} plots the same as Figure \ref{fig:sep} but as a function of $m_{\star b}\equiv \log\left[M_{\star b}/(\munit)\right]$.
As can be seen, we still fail to find any indication of the mass dependent spin transition at fixed morphology. 
Note that a significant difference exists in their alignment tendency between the $Sab$ and $Scd$ galaxies even at the same bulge masses, which implies 
the transition from parallel to perpendicular spin alignments may not be triggered mainly by the growth of central bulges but by some evolutionary process through 
which the late-type spirals are segregated from the early-type ones. 

\section{Summary and Discussion}\label{sec:con}

We have observationally detected a statistically significant signal of mass dependent galaxy spin transition by investigating the alignments between the spin axes of the 
spiral galaxies located at the void outskirts and the directions toward the centers of the nearest voids as a function of the galaxy stellar mass. 
For this investigation, we have considered only the SDSS spiral galaxies whose three dimensional spin axes can be calibrated from the photometric position 
angles and axial ratios up to the two-fold ambiguity via the scheme based on the circular thin disk approximation \citep{LE07,lee11}. 
We have also analyzed the most updated catalogs of the voids identified by \citet{void-catalog} via three different algorithms (VIDE, REVOLVER, and VoidFinder), 
assuming that the direction from a given galaxy to its nearest void center is parallel to the direction of maximum matter compression in the galaxy neighborhood. 

It has been found that the spiral galaxy spins indeed gradually transit from parallel to perpendicular orientations with respect to the directions of maximum 
matter compression as the galaxy stellar mass increases, regardless of the void-identification algorithms. However, the coexistence of the statistically significant signals 
($>3\sigma$) of the parallel and perpendicular spin alignments in the smallest and largest mass bins, respectively, has been detected only for the VoidFinder case, where 
$1\sigma$ has been obtained from the $10,000$ resamples created by random shuffling of the galaxy positions. For this case, the threshold stellar mass, $M_{\rm th,\star}$, 
at which the galaxy spin transition occurs has been efficiently constrained to the range of $9.51\le\log\left[M_{\rm th,\star}/(\munit)\right]\le10.3$ (spin transition zone) 
for both of the Planck and WMAP5 cosmologies by employing the KS-test based methodology developed by \citet{LL20}. 

At a fixed morphological type, however, has been witnessed disappearance of the spin transition signals.   The $Sab$ galaxies exhibit only {\it perpendicular} spin alignments 
in an almost mass independent manner, in contrast to the $Scd$ galaxies whose spins yield only insignificant {\it parallel} alignments in the entire stellar mass range. Even when 
the alignments are measured as a function of the bulge masses, a similar result has been found, which implies that disappearance of the spin transition signals at fixed 
morphology cannot be ascribed to the difference between the $Sab$ and $Scd$ galaxies in the bulge masses. 

Yet, the high statistical significance of the spin transition signal we have been able to attain here is owing to the large size of our sample, which has in turn been possible 
to obtain because we have determined the spiral galaxy spins by using the {\it photometric} position angles and axial ratios.
Although a higher accuracy in the determination of the spiral galaxy spins could have be achieved by using more reliable kinematic position angles rather than the 
photometric counterparts as in \citet{kra-etal21} and \citet{bar-etal22}, we have chosen the latter to avoid a poor number statistics, since information on kinematic position angles 
from well resolved data is available only for a limited number of the spiral galaxies. 
Furthermore, for the case of the spiral galaxies located at the outskirts of cosmic voids characterized by the lowest environmental densities, 
the offsets between their photometric and kinematic position angles were observationally known to be as low as $\le 10^{\circ}$ 
\citep[e.g.,][]{BS03,bar-etal14,bar-etal15,gra-etal18}, which justifies our choice in some degree. 
 
Providing an observational evidence for the occurrence of the galaxy spin transition phenomenon in the {\it non-filamentary environments}, our results bear three 
physical implications. First, the filamentary merging of the galaxies may not be the most crucial mechanism that drives the spiral galaxies to transit their spin orientations 
with respect to the directions of maximum matter compression. Second, the most decisive factor that triggers the occurrence of the spin transition should be linked 
not with the growth of total or bulge stellar masses of the spiral galaxies but with some other mechanism responsible for the morphology segregation.  
Third, it becomes quite feasible to use the observed spin transition zone as a discriminator of cosmology, since we have found no difference 
in the spin transition zone between the Planck and WMAP5 $\lcdm$ cosmologies, while it was previously found that it significantly differs between the standard 
$\lcdm$ and non-standard cosmologies \citep{lee-etal20,LL20}. 

It is also worth listing two crucial questions raised by the current work. The first one is why the mass-dependent spin transition threshold for the spiral galaxies on the void 
surfaces can be efficiently constrained only when the voids are identified by the VoidFinder algorithm. The second one is why the mass-dependent spin transition 
signal disappears at fixed morphology, and what factor plays the most decisive role for the spiral galaxy spin transition, if it is not the growth of total/bulge stellar masses. 
Our future work will be in the direction of finding answers to these questions by performing a comprehensive numerical and observational analysis.
 
\acknowledgments

Funding for the SDSS and SDSS-II has been provided by the Alfred P. Sloan Foundation, 
the Participating Institutions, the National Science Foundation, the U.S. Department of 
Energy, the National Aeronautics and Space Administration, the Japanese Monbukagakusho, 
the Max Planck Society, and the Higher Education Funding Council for England. The 
SDSS Web Site is http://www.sdss.org/. 

The SDSS is managed by the Astrophysical Research Consortium for the Participating 
Institutions. The Participating Institutions are the American Museum of Natural History, 
Astrophysical Institute Potsdam, University of Basel, University of Cambridge, Case 
Western Reserve University, University of Chicago, Drexel University, Fermi lab, the 
Institute for Advanced Study, the Japan Participation Group, Johns Hopkins University, 
the Joint Institute for Nuclear Astrophysics, the Kavli Institute for Particle 
Astrophysics and Cosmology, the Korean Scientist Group, the Chinese Academy of Sciences 
(LAMOST), Los Alamos National Laboratory, the Max-Planck-Institute for Astronomy (MPIA), 
the Max-Planck-Institute for Astrophysics (MPA), New Mexico State University, Ohio State 
University, University of Pittsburgh, University of Portsmouth, Princeton University, 
the United States Naval Observatory, and the University of Washington. 

We are very grateful to an anonymous referee whose valuable comments helped us significantly improve the original manuscript. 
JL acknowledges the support by Basic Science Research Program through the National Research Foundation (NRF) of Korea funded 
by the Ministry of Education (No.2019R1A2C1083855). JSM acknowledges the support by the NRF of Korea grant funded by the Korean 
government (MEST) (No. 2019R1A6A1A10073437).

\clearpage
\clearpage
\clearpage
\clearpage
\clearpage
\clearpage
\begin{deluxetable}{ccccccccc}
\tablewidth{0pt}
\tablecaption{Numbers of the selected spirals belonging to the six mass-selected samples}
\setlength{\tabcolsep}{3mm}
\tablehead{Type & Model & Algorithm & $N_{g}(S_{1})$ & $N_{g}(S_{2})$ & $N_{g}(S_{3})$ & $N_{g}(S_{4})$ & $N_{g}(S_{5})$ & $N_{g}(S_{6})$}
\startdata
& & VIDE &  $14999$ & $14399$ & $14512$ & $14593$ & $14700$ & $14769$ \\
All & Planck & REVOLVER &  $13425$ & $12434$ & $12431$ & $12468$ & $12349$ & $12508$ \\
& & VoidFinder & $22501$ & $22495$ & $22524$ & $22494$ & $22460$ & $22580$ \\
\hline
& & VIDE &  $14998$ & $14383$ & $14475$ & $14558$ & $14646$ & $14749$ \\
All & WMAP5 & REVOLVER & $13437$ & $12453$ & $12431$ & $12480$ & $12355$ & $12468$ \\
& & VoidFinder &  $22501$ & $22495$ & $22524$ & $22494$ & $22460$ & $22580$ \\
\hline
& & VIDE &  $8805$ & $8637$ & $8620$ & $8667$ & $8848$ & $8693$ \\
$Sab$ & Planck & REVOLVER &  $7846$ & $7507$ & $7373$ & $7353$ & $7429$ & $7393$ \\
& & VoidFinder & $8881$ & $8967$ & $9117$ & $9309$ & $9392$ & $9371$ \\
\hline
& & VIDE &  $8789$ & $8615$ & $8587$ & $8644$ & $8795$ & $8688$ \\
$Sab$ & WMAP5 & REVOLVER & $7836$ & $7504$ & $7361$ & $7366$ & $7393$ & $7389$ \\
& & VoidFinder &  $8981$ & $9003$ & $9111$ & $9303$ & $9349$ & $9357$ \\
\hline
& & VIDE &  $6275$ & $5755$ & $5906$ & $5907$ & $5880$ & $5979$ \\
$Scd$ & Planck & REVOLVER &  $5599$ & $5108$ & $4985$ & $5030$ & $4984$ & $5008$ \\
& & VoidFinder & $5897$ & $5851$ & $5951$ & $6019$ & $6211$ & $6208$ \\
\hline
& & VIDE &  $6274$ & $5763$ & $5900$ & $5898$ & $5869$ & $5987$ \\
$Scd$ & WMAP5 & REVOLVER & $5601$ & $5144$ & $4992$ & $5031$ & $4986$ & $5021$ \\
& & VoidFinder &  $5909$ & $5932$ & $5993$ & $6033$ & $6190$ & $6199$ \\
\enddata
\label{tab:ng}
\end{deluxetable}

\end{document}